\documentclass[aps,preprint]{revtex4}%
\usepackage{amsfonts}%
\usepackage{amsmath}%
\usepackage{amssymb}%
\usepackage{graphicx}

\begin{document}
\title{Nonequilibrium superconductivity in Y$_{1-x}$Pr$_{x}$Ba$_{2}$Cu$_{3}$O$_{7}$
thin films}
\author{R. D. Averitt, V. K. Thorsm\o lle, Q. X. Jia, S.A. Trugman, A.J Taylor}
\affiliation{Los Alamos National Lab, MST-10 Condensed Matter and Thermal Physics, MS K764,
Los Alamos, NM 87544}

\begin{abstract}
We have measured the picosecond conductivity dynamics in Y$_{1-x}$Pr$_{x}%
$Ba$_{2}$Cu$_{3}$O$_{7}$ thin films from 0.3-2.0 terahertz. Our experimental
technique measures the complex conductivity $\sigma_{re}\left(  \omega\right)
$ + i$\sigma_{im}\left(  \omega\right)  $ permitting the simultaneous
observation of superconducting pair and quasiparticle dynamics. We emphasize
aspects of the conductivity dynamics which extend our previous results
\cite{averitt01PRB}. In particular, the recovery of $\sigma_{re}$ is faster
than $\sigma_{im}$, and the recovery of $\sigma_{im}$ decreases with
increasing frequency. This suggests another carrier relaxation pathway, in
addition to superconducting pair recovery, following optical excitation. 

\end{abstract}
\maketitle

Terahertz time-domain spectroscopy (THz-TDS) is an ultrafast optical technique
that has found wide application in the study of many systems having
far-infrared excitations. In the context of correlated electron materials,
THz-TDS has been successfully applied to study high-T$_{c}$ superconductors
and, more recently, materials such as the ferromagnetic metal SrRuO$_{3}$
\cite{corson00PRL,dodgeetal00PRL}. We can expect an increase in the use of
THz-TDS to study a variety of other correlated electron materials given its
unique ability to directly and easily measure $\sigma_{re}$($\omega$) +
i$\sigma_{im}$($\omega$) from $\sim$50 GHz to several THz.

Importantly, the freely propagating THz pulses generated in THz-TDS are
temporally coherent with the generating optical pulses - this permits
measurement of the THz conductivity with picosecond resolution following
optical excitation of the sample. Several groups have been developing this
technique, termed time-resolved THz spectroscopy (TRTS), to study various
systems including photogenerated electrons in liquids such as hexane, or
semiconductors such as GaAs \cite{knoesel01PRL,beard00PRB}. Our work has
focused on using TRTS to study high-T$_{c}$ superconductors and colossal
magnetoresistance manganites
\cite{averitt01PRB,averittetal00JOSAB,averittetal01PRL}. Here we present our
most recent measurements on Y$_{1-x}$Pr$_{x}$Ba$_{2}$Cu$_{3}$O$_{7}$ thin films.

Figure 1(a) and (b) show the conductivity at 60K and 95K respectively (T$_{c}$
= 89K) for the near optimally doped film. The phenomenological two-fluid model
fits the data quite well (shown as a dashed line in Fig. 1(a)) below T$_{c}$
where the imaginary conductivity is dominated by the 1/$\omega$ dependence of
the superfluid \cite{averitt01PRB,brorson96JOSAB}. Above T$_{c}$, a standard
Drude model fits the data (dashed line, Fig. 1(b)). Upon optical excitation,
there is a decrease in the imaginary conductivity due to superconducting pair
breaking with a corresponding increase in the real component (not shown). The
induced change in $\sigma_{im}(\omega)$ is shown at 60K in figure 1(c). There
is a decrease that rapidly recovers on a ps timescale that is due in large
part to superconducting pair reformation. Figure 1(d) shows the induced change
in $\sigma_{im}(\omega)$ at 95 K (above T$_{c}$) which is due quasiparticle relaxation.

The dynamics can be followed by plotting the induced change in the
conductivity as a function of time at a specified frequency. The induced
change in the imaginary conductivity (60K) is shown in Figure 2. With
increasing frequency, the lifetime decreases (see inset). In the limit of zero
quasiparticle fraction, this induced change would be solely due to
superconducting pair recovery. However, there are quasiparticles present (at
60K the initial quasiparticle fraction is $\sim$ 40 $\%$) so the quasiparticle
fraction makes a nonnegligible contribution to $\sigma_{im}(\omega)$. At
higher frequencies this fraction becomes increasingly important since the
superconducting pair fraction response goes as 1/$\omega$. This offers a
potential explanation for the decrease in the lifetime of $\sigma_{im}%
(\omega)$ with increasing frequency: at low frequencies $\sigma_{im}(\omega)$
is dominated by the superconducting pair recovery, but at higher frequencies
the relaxation is increasingly influenced by an additional relaxation pathway
associated with the quasiparticles. This is further supported in that the
lifetime of $\sigma(\omega)_{re}$ is quite short ($\sim$ 1.5 ps independent of
frequency). Speculating on the origin of this additional relaxation pathway,
it could be due to the relaxation of the excited quasiparticles into the nodes
of the superconducting gap along k$_{x}$ = k$_{y}$. Since this process is
faster than the superconducting pair recovery, this suggests that the excited
quasiparticles relax into the nodes of the gap followed by pair recovery.

Finally, the results of the superconducting pair recovery time as a function
of temperature are consistent with our previous results \cite{averitt01PRB}.
For optimal doping, the lifetime at 20 K is about 1.5 ps (at 1.0 THz)
increasing to 3.0 ps near T$_{c}$. Above T$_{c}$, the lifetime of $\sigma
_{im}(\omega)$ (which is no longer a measure of the superconducting recovery
time, but rather the initial quasiparticle cooling) drops to 1.5 ps (this is
also consistent with the discussion in the previous paragraph). In contrast,
in the x=0.3 films the lifetime is $\sim$3.5 ps independent of temperature
even above T$_{c}$. This lifetime is the same as that measured in our
YBa$_{2}$Cu$_{3}$O$_{6.5}$ films. These results suggest that for the optimally
doped films, the dynamics are influenced by the closing of the superconducting
gap, and that for the underdoped films, the pseudogap plays a role in
determining a observed dynamics.

Research supported by the Los Alamos Laboratory Directed Research and
Development Program, Department of Energy.

\newpage Figure 1: (a)-(b) Conductivity at 60K and 95K, respectively for
YBa$_{2}$Cu$_{3}$O$_{7}$. The imaginary conductivity is plotted with the real
conductivity in the insets. The data is fit using a two-fluid model which
reduces to the Drude model above T$_{c}$ \cite{averitt01PRB,brorson96JOSAB}.
The thick solid line is the experimental data. The dashed lines are the
overall fit to the data. In (a), the dotted line is the superconducting pair
component and the thin solid line is the quasiparticle component. (c)-(d)
Optically induced changes in the imaginary conductivity at 60 and 95 K at an
excitation fluence of 12 $\mu J/cm^{2}.$

\bigskip

Figure 2: Normalized induced change in the imaginary conductivity (60 K) as a
function of time at various frequencies. The curves are displaced vertically
for clarity. The solid lines are fits to the function y = a$\times
$exp(-t/$\tau_{\sigma}$) + b. The inset shows the measured lifetime as a
function of frequency.

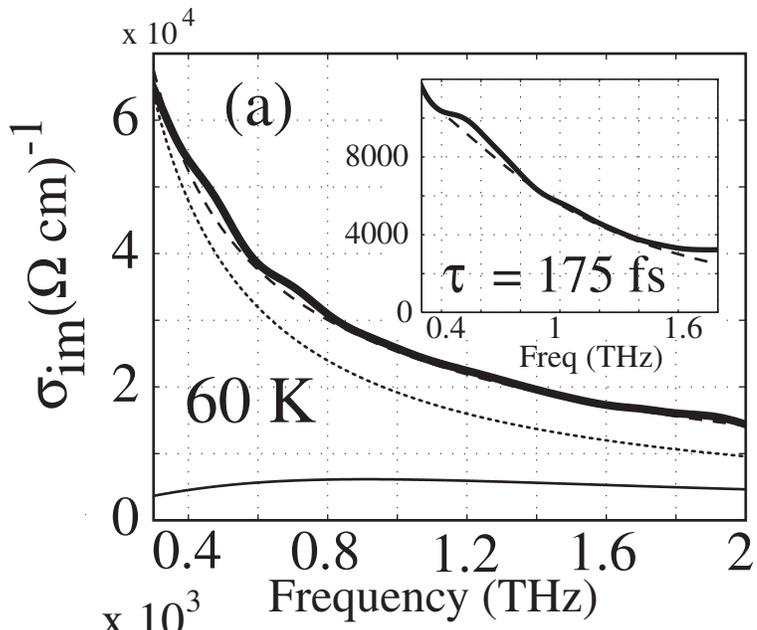
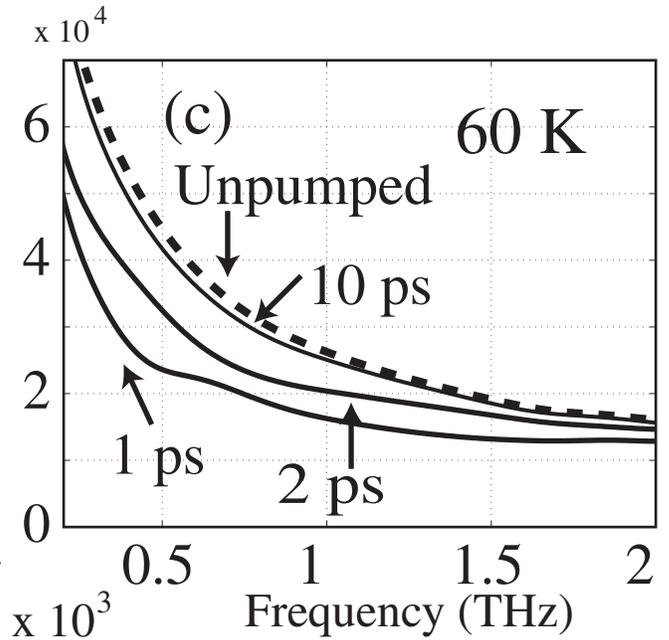
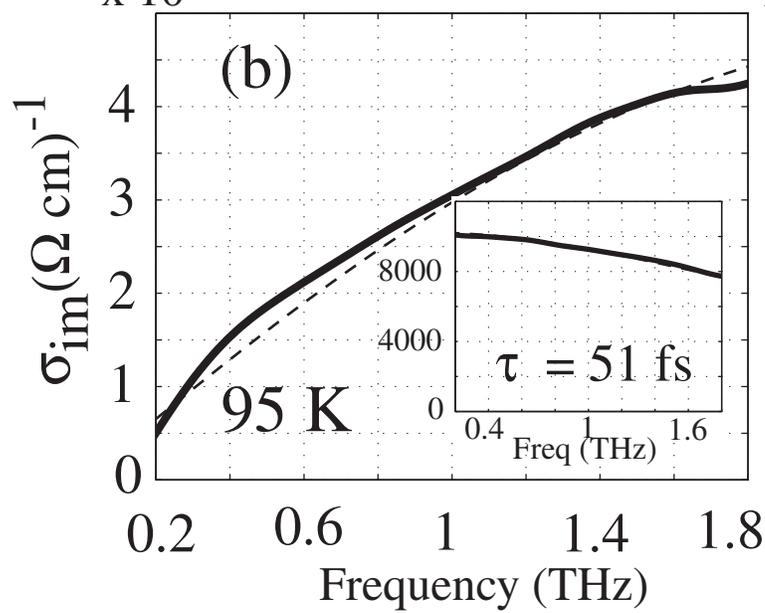
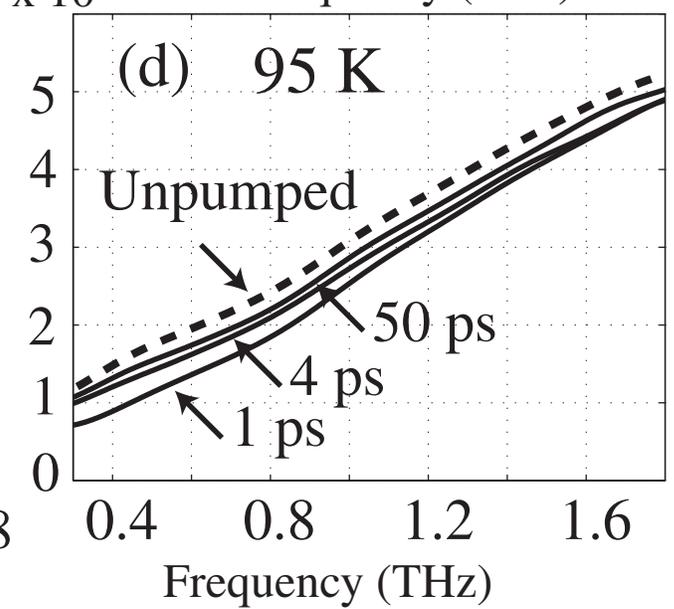
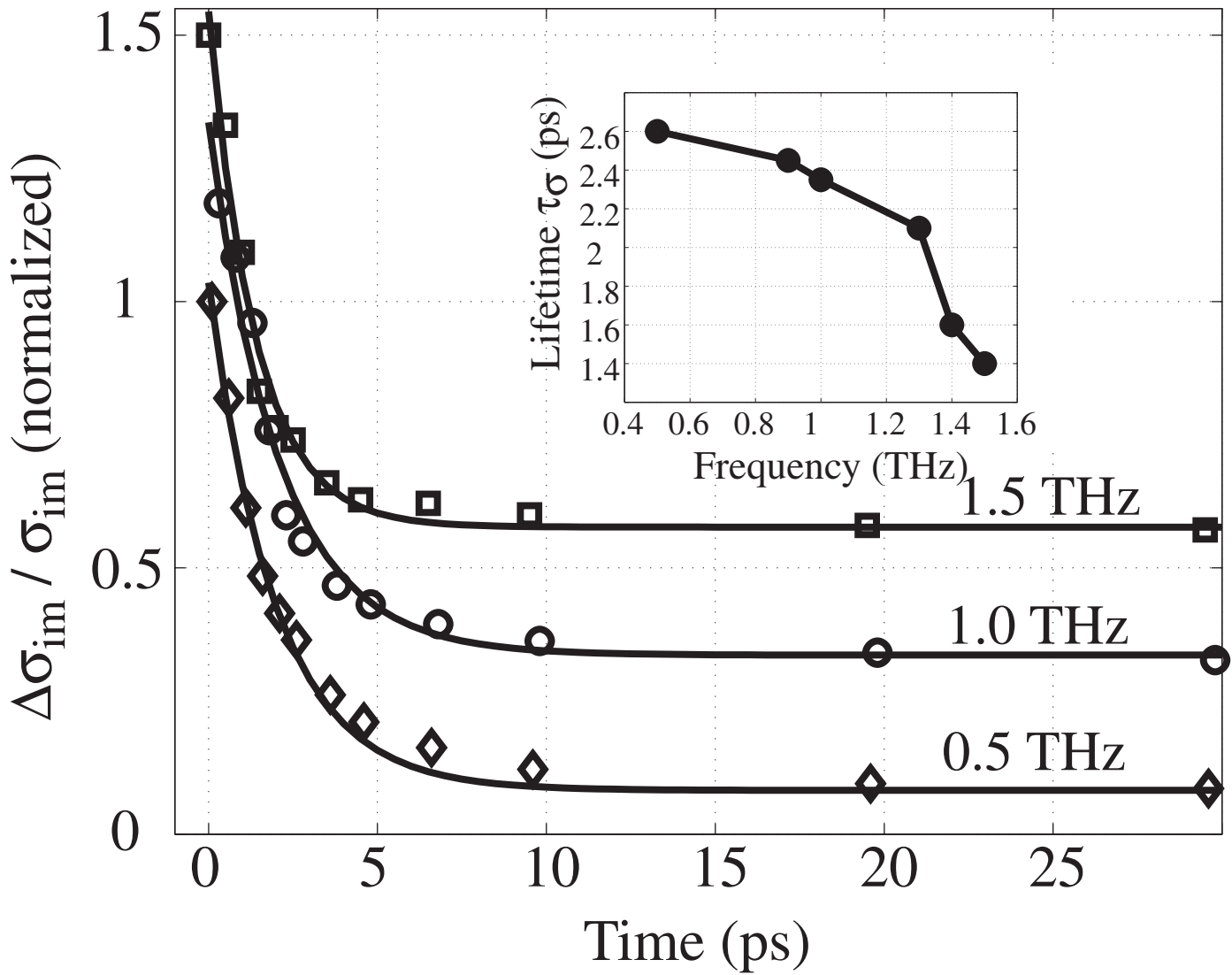
\end{document}